\newcommand{\eps}{\epsilon}
\newcommand{\en}{\epsilon_0}
\newcommand{\sx}{\sigma_x}

\newcommand{\sz}{\sigma_z}
\newcommand{\dbz}{\Delta B_z}

\newcommand{\W}{\Omega}

\newcommand{\bra}[1]{\left| #1 \right \rangle}
\newcommand{\ta}[1]{\left \langle #1 \right \rangle}

\documentclass[%
reprint,amsmath,amssymb,aps,prl,superscriptaddress,
]{revtex4-1}

\usepackage{graphicx}
\usepackage{dcolumn}
\usepackage{bm}
\usepackage{epstopdf}

\begin{document}

\title{High-fidelity entangling gate for double-quantum-dot spin qubits}

\author{John M. Nichol}
\thanks{These authors contributed equally.}

\author{Lucas A. Orona}
\thanks{These authors contributed equally.}

\author{Shannon P. Harvey}
\affiliation{Department of Physics, Harvard University, Cambridge, MA, 02138, USA}

\author{Saeed Fallahi}
\affiliation{Department of Physics and Astronomy, Purdue University, West Lafayette, IN, 47907 USA}
\affiliation{Birck Nanotechnology Center, Purdue University, West Lafayette, IN, 47907 USA}

\author{Geoffrey C. Gardner}
\affiliation{Birck Nanotechnology Center, Purdue University, West Lafayette, IN, 47907 USA}
\affiliation{School of Materials Engineering, Purdue University, West Lafayette, IN, 47907 USA}

\author{Michael J. Manfra}
\affiliation{Department of Physics and Astronomy, Purdue University, West Lafayette, IN, 47907 USA}
\affiliation{Birck Nanotechnology Center, Purdue University, West Lafayette, IN, 47907 USA}
\affiliation{School of Materials Engineering, Purdue University, West Lafayette, IN, 47907 USA}
\affiliation{School of Electrical and Computer Engineering, Purdue University, West Lafayette, IN, 47907 USA}

\author{Amir Yacoby}
\email{yacoby@physics.harvard.edu}

\affiliation{Department of Physics, Harvard University, Cambridge, MA, 02138, USA}

\begin{abstract}
Electron spins in semiconductors are promising qubits~\cite{Loss1998,Petta2005,Kim2014,Koppens2006,Eng2014,Pioro-Ladriere2008,Muhonen2014} because their long coherence times enable nearly $10^9$ coherent quantum gate operations~\cite{Veldhorst2014}. However, developing a scalable high-fidelity two-qubit gate remains challenging. Here, we demonstrate an entangling gate between two double-quantum-dot spin qubits in GaAs~\cite{Petta2005} by using a magnetic field gradient between the two dots~\cite{Foletti2009} in each qubit to suppress decoherence due to charge noise. When the magnetic gradient dominates the voltage-controlled exchange interaction between electrons, qubit coherence times increase by an order of magnitude. Using randomized benchmarking and self-consistent quantum measurement, state, and process tomography, we measure single-qubit gate fidelities of approximately $99 \%$ and an entangling gate fidelity of $90\%$. In the future, operating double quantum dot spin qubits with large gradients in nuclear-spin-free materials, such as Si, should enable a two-qubit gate fidelity surpassing the threshold for fault-tolerant quantum information processing.
\end{abstract}


\maketitle
The quantum phase coherence of isolated spins in semiconductors can persist for long times, reaching tens of milliseconds for electron spins~\cite{Veldhorst2014} and tens of minutes for nuclear spins~\cite{Saeedi2015}. Such long coherence times enable single-qubit gate fidelities exceeding the threshold for fault-tolerant quantum computing~\cite{Veldhorst2014} and make spins promising qubits. However, entangling spins is difficult because spin-spin interactions are weak. For electrons, this challenge can be met by exploiting the charge of the electron for electric-dipole~\cite{Shulman2012} or gate-controlled exchange coupling~\cite{Petta2005} between spins. In these methods, however, the qubit energy depends on electric fields, and charge noise in the host material limits single-qubit coherence~\cite{Dial2013}. Charge noise also affects other qubit platforms. For example, heating due to charge noise is a limiting factor in the coherence of trapped ion qubits~\cite{Brownnutt2015}, and the transmon superconducting qubit was designed to suppress noise from charge fluctuations in superconducting islands~\cite{Houck2009}. Strategies such as composite pulses~\cite{Yang2016,Cerfontaine2016}, dynamical decoupling~\cite{Dial2013}, and sweet-spot operation~\cite{Reed2016,Martins2016} have been developed to mitigate the effects of charge noise. 

In this work, we present a technique to suppress decoherence caused by charge noise. The key idea is to apply a large transverse qubit energy splitting that does not depend on electric fields and therefore suppresses the effects of charge fluctuations. We implement this scheme with two singlet-triplet qubits, each of which consists of two electrons in a double-quantum-dot~\cite{Petta2005}. In each qubit, the voltage-controlled exchange interaction $J(\eps)$, where $\eps$ represents the gate voltage, splits the singlet $\bra{S}=(\bra{\uparrow \downarrow}- \bra{\downarrow \uparrow})/\sqrt{2}$ and triplet  $\bra{T_0}=(\bra{\uparrow \downarrow}+ \bra{\downarrow \uparrow})/\sqrt{2}$ states in energy ~\cite{Petta2005}, where the left(right) arrow indicates the spin of the left(right) electron. A magnetic gradient $\dbz$ between the two dots lifts the degeneracy between $\bra{\uparrow \downarrow}$ and $\bra{\downarrow \uparrow}$. These two mechanisms enable universal quantum control of singlet-triplet qubits~\cite{Foletti2009}. Until now, two-qubit gates for singlet-triplet qubits have operated with $J(\eps)\gg\dbz$, and charge noise is the limiting factor in two-qubit gate fidelities~\cite{Petta2005,Shulman2012}. However, if $\dbz \gg J(\eps)$, the total qubit energy splitting is $\Omega(\eps) = \sqrt{\dbz^2+J(\eps)^2}\approx \dbz+\frac{J(\eps)^2}{2\dbz}$, and the qubit sensitivity to charge noise $\W'(\eps)=\frac{J(\eps)}{\dbz}J'(\eps)$ is reduced by a factor of $\frac{J(\eps)}{\dbz}$, effectively mitigating decoherence due to charge noise. 

Intense magnetic field gradients in spin qubits can be created with micromagnets~\cite{Pioro-Ladriere2008,Takeda2016,Wu2014}. In GaAs quantum dots, strong magnetic gradients can also be generated via the hyperfine interaction between the electron and Ga and As nuclear spins in the semiconductor~\cite{Foletti2009,Bluhm2010,Shulman2014,Nichol2015}. Coherence times for qubit rotations around hyperfine gradients can approach one millisecond~\cite{Malinowski2016}, which is significantly longer than typical exchange coherence times~\cite{Dial2013}. Here, we show that when the magnetic gradient in a GaAs singlet-triplet qubit dominates the electrically-controlled exchange interaction, coherence times increase by an order of magnitude.  Through both standard and interleaved randomized benchmarking, we measure average single qubit gate fidelities of approximately $99 \%$. At the same time, this approach maintains a large interaction between adjacent capacitively coupled qubits. We use self-consistent two-qubit state- and measurement tomography to measure a Bell state with a maximum fidelity of $93\%$. Full process tomography involving 256 tomographic measurements of the two-qubit operation yields an entangling gate fidelity of approximately $90\%$, consistent with theoretical simulations. In materials without nuclear spins such as silicon, even higher gate fidelities should be possible.

\begin{figure}
\includegraphics{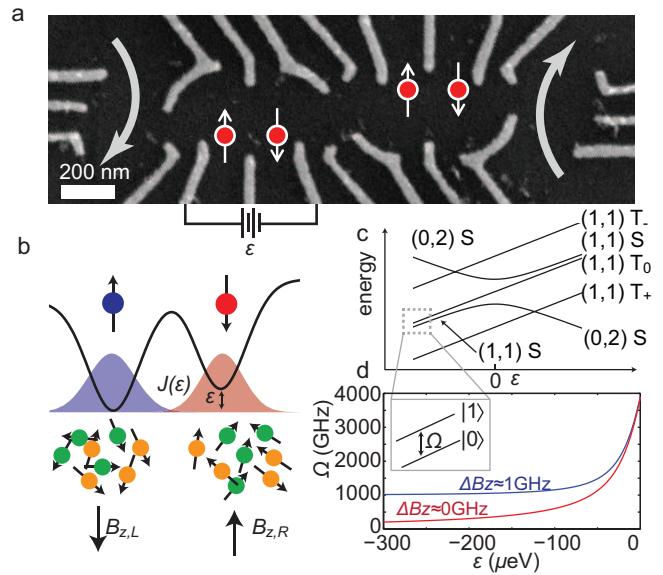}
\caption{\label{apparatus} Experimental setup. (a) Scanning electron micrograph of a two-qubit device identical to the one used in this work. Red circles indicate approximate positions of electrons in the double-well potentials created by metal depletion gates (gray). Arrows indicate the positions of sensor quantum dots. A voltage difference $\eps$ applied to plunger gates adjusts the exchange interaction.  (b) The gate-contolled wavefunction overlap between electron spins produces the exchange interaction $J(\eps)$. Each electron also interacts with a large number of Ga and As nuclear spins (green and orange circles) via the hyperfine interaction, leading to a difference in the logitudinal magnetic gradient between the dots $\dbz=B_{z,L}-B_{z,R}$. (c) Energy level diagram showing the two-electron spin states of a double quantum dot. We operate the qubit with $\eps<0$ and $J(\eps) \ll \dbz$, as indicated with the dashed gray box. (d) Calculated qubit energy splitting $\Omega(\eps)$ for the two cases when $\dbz=0$ and $\dbz \approx 1$ GHz. When $\dbz$ is large, the qubit splitting does not depend on $\eps$ and is insensitive to electric fields.}
\end{figure}

We use two singlet-triplet qubits~\cite{Petta2005}, created in gate-defined double quantum dots similar to those of refs.~\cite{Dial2013,Shulman2012} in a GaAs/AlGaAs heterostructure [Fig.~\ref{apparatus}(a)]. Each double quantum dot contains two electrons. The Hamiltonian for each qubit is $H(\eps)=J(\eps) \sigma_z + \Delta B_z \sigma_x$, in the $\{\bra{S}, \bra{T_0}\}$ basis. $J(\eps)$, the exchange interaction between the two spins, depends on $\eps$, the difference in electrochemical potential between the dots [Fig.~\ref{apparatus}(b)]. $\dbz$, the difference in longitudinal magnetic field between the two dots, results from the wavefunction overlap between each electron and the Ga and As nuclear spins in the heterostructure.  Although the nuclear spins are unpolarized in thermal equilibrium, $\dbz$ can be measured and stabilized up to several hundred mT using feedback~\cite{Foletti2009,Bluhm2010,Shulman2014}.  

The two adjacent qubits are capacitively coupled, and the interaction Hamiltonian $H_{int}= J_{12} \sigma_z\otimes\sigma_z$, where $J_{12}\propto J'_1(\eps_1)J'_2(\eps_2)$~\cite{Taylor2005,Shulman2012}, and the subscripts refer to the different qubits. For the values of $\eps$ used here, we empirically find that $J'(\eps) \propto J(\eps)$. This requires that $J(\eps)>0$ to maintain nonzero interqubit coupling.

Figure~\ref{apparatus}(c) shows the energy level diagram of the two-electron spin states in a double quantum dot. The qubit states are the $\bra{S}$ and $\bra{T_0}$ levels in the regime where $\dbz \gg J(\eps)$ [Fig.~\ref{apparatus}(c-d)]. Through dynamic nuclear polarization and feedback, we set $g^* \mu_B \dbz/h \approx $1 GHz in all experiments~\cite{Foletti2009,Bluhm2010}. Here $g^* = -0.44$ is the effective electron g-factor in GaAs, $\mu_B$ is the Bohr magneton, and $h$ is Planck's constant. $\dbz$ is stabilized to within 3 MHz, corresponding to an inhomogeneously broadened coherence time $T_2^* \approx 100$ ns.  We initialize  the $\bra{0}$ state through electron exchange with the leads when $\eps \gg 0$, where $\bra{S}$ is the ground state of the double dot. Then we adiabatically ramp to $\eps=\en < 0$, where $100~\textnormal{MHz} <J(\en)/2 \pi < 300 ~\textnormal{MHz} < \dbz$. We measure the qubit state via electron exchange with the leads in a new technique (see Supplementary Information), which is compatible with large magnetic gradients~\cite{Barthel2012}. 

\begin{figure}
\includegraphics{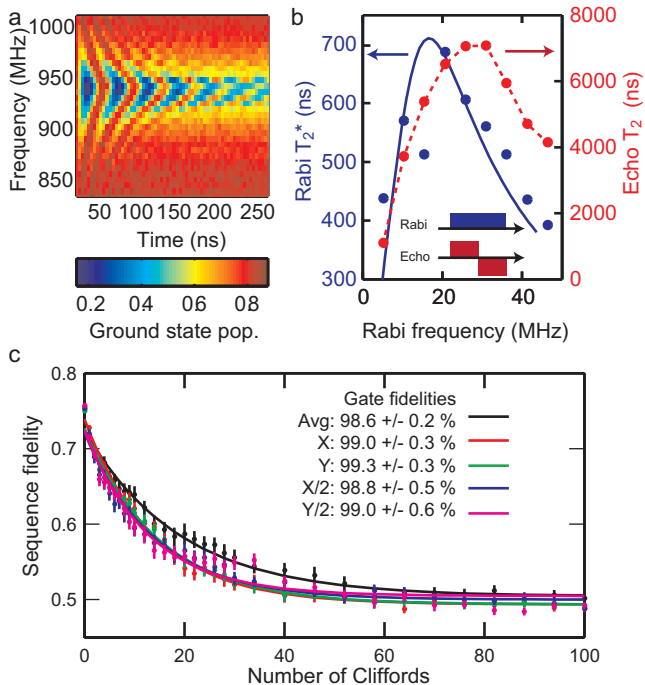}
\caption{\label{singleQubit} Single-qubit operations. (a) Time-varying voltage pulses resonant with the qubit splitting $\Omega$ induce Rabi oscillations. (b) Coherence times of driven Rabi oscillations (blue) and rotary echo (red) vs. Rabi drive strength. The solid blue line is a theoretical curve taking into account the measured charge and hyperfine noise levels in our qubit. The data agree with the model. The dashed red line between data points is a guide to the eye. At low drive strengths, hyperfine fluctuations limit the coherence time, and at large drive strengths, charge-noise-induced fluctuations in the Rabi frequency limit the coherence. (c) Randomized benchmarking yields an average gate fidelity of $98.6 \pm 0.2 \%$. Error bars are statistical uncertainties.}
\end{figure}

We drive qubit rotations by adding an oscillating voltage to the plunger gates, such that the total voltage $\eps(t)=\en+\eps_1 \cos(\Omega t)$.  For $\eps_1 \ll \eps_0$, $J(t) \approx J(\en)+2j \cos(\Omega t)$, where $j=\frac{\eps_1}{2}J'(\en)$ is the Rabi frequency. When the oscillation frequency matches the total qubit splitting $\Omega = \sqrt{\dbz^2+J^2(\en)}$, the time varying component of $J(t)$ drives qubit transitions [Fig. ~\ref{singleQubit} (a)]~\cite{Shulman2014}. In this regime, $\Omega \approx \dbz$ is analogous to the external magnetic field for a single spin-1/2, while the time varying component of $J(t)$ is analogous to a perpendicular oscillating magnetic field, which drives transitions. We emphasize that when $\dbz \gg J(\en)$, $\Omega(\en) \approx \dbz+\frac{J(\en)^2}{2\dbz}$, and the sensitivity to charge noise $\W'(\en)=\frac{J(\en)}{\dbz}J'(\en)$ is smaller by a factor of $\frac{J(\en)}{\dbz}$ compared to the case where $\dbz \ll J(\en)$ [Fig.~\ref{apparatus}(d)]. However, a key requirement of this technique is that $J(\eps_0)>0$, in order to maintain $J'(\en)>0$ for single-qubit control and two-qubit coupling.

Large magnetic gradients can therefore completely suppress dephasing due to charge noise, although relaxation caused by charge noise at the qubit frequency $\dbz$ still limits the coherence. In our case, however, nuclear spin noise causes the magnetic gradient to fluctuate. To suppress the effects of hyperfine fluctuations, we apply a strong rf drive to the qubit, causing Rabi oscillations. In the reference frame rotating around the qubit splitting $\W \sigma_x$, the Hamiltonian is $H_{rot}=j\sigma_z +\delta \W \sigma_x$, where $j$ is the Rabi frequency, and $\delta \W$ is a fluctuation in the magnetic gradient. When $j \gg \delta \W$, the qubit splitting in the rotating frame $\W_{rot} \approx j+\frac{\delta \W^2}{2j}$ is first-order insensitive to fluctuations in the magnetic gradient.

Figure~\ref{singleQubit}(b) shows the coherence time of driven Rabi oscillations as a function of drive strength for $J(\en)/(2 \pi) = 220$ MHz. The maximum coherence time ($\approx 700$ ns) is an order of magnitude larger than that for oscillations around a static exchange splitting with the same $J(\en)$ ($\approx 80$ ns). However, the quality factor of Rabi oscillations is the same as for static exchange oscillations~\cite{Dial2013}, because low-frequency charge noise limits the coherence time in both cases. But because $j \ll J(\eps_0)$, the Rabi coherence time is much longer. It is this improvement in coherence that allows increased two-qubit gate fidelities, as described below. Reversing the phase of the drive halfway through the evolution to perform a rotary echo extends the coherence time by an additional factor of 10 [Fig.~\ref{singleQubit}(b)]. Rotating-frame echo coherence times are also an order of magnitude longer than static exchange echo~\cite{Dial2013} dephasing times measured in this device. 

As the amplitude of the oscillating voltage $\eps_1$ increases, both the Rabi and echo coherence times reach a maximum [Fig.~\ref{singleQubit}(b)]. At low drive strengths, hyperfine fluctuations in the detuning limit the coherence. At large drive strengths, charge-noise-induced fluctuations in $J'(\eps)$, which cause the Rabi rates to fluctuate in time, limit the coherence. The observed behavior agrees well with a theoretical simulation based on measured noise levels in our qubit [Fig.~\ref{singleQubit}(b)] (see Supplementary Information). The simulation correctly predicts the maximum coherence time and corresponding Rabi frequency. Using randomized benchmarking~\cite{Knill2007}, we find an average gate fidelity of $98.6 \pm 0.2 \%$ [Fig.~\ref{singleQubit}(c)]. Interleaved randomized benchmarking~\cite{Magesan2012} reveals individual gate fidelities close to the measured average fidelity.  Gate fidelities are likely coherence limited as a result of slow electric-field or hyperfine fluctuations. Given the observed quality factor of Rabi oscillations, which is approximately 5, [Fig.~\ref{singleQubit}(a)], we would expect roughly 10 coherent $\pi$ rotations within the coherence time. Assuming Gaussian decay due to low-frequency noise, the fidelity of a $\pi$-gate should be approximately $e^{-(1/10)^2}=0.99$. Because hyperfine or charge fluctuations are slow compared with gate times ($\approx20~$ns), errors are likely correlated~\cite{Ball2016}, as is the case for most spin qubits. Suppressed low-frequency charge noise or composite pulses~\cite{Yang2016,Cerfontaine2016} would improve gate fidelities.

Next, we take advantage of the long coherence times in the $\dbz$-dominated regime to perform a high-fidelity two-qubit entangling gate. In the lab frame, $H\approx \W_1 \sx \otimes I + \W_2 I \otimes \sx + J_{12}\sz\otimes\sz$, where $I$ is the identity operator. The single-qubit terms in the Hamiltonian do not commute with the interaction term, and the single-qubit rotations cancel the interaction except when $\W_1=\W_2$ (see Supplementary Information). In this case, the interaction in the rotating frame is
\begin{align}
H_{int} \approx \frac{J_{12}}{2} \sz \otimes \sz \cos(\phi_1-\phi_2) \label{eq:Interaction}.
\end{align} 
Here $J_{12} \propto J'_1(\eps_1)J'_2(\eps_2)$ is the interaction energy, and $\phi_i$ is the phase of the rf drive on each qubit. When  $\W_1 = \W_2$ and $\phi_1=\phi_2$, the single-qubit rotations constructively interfere, and the interaction is the same as in the lab frame, up to a factor of $1/2$. The order-of-magnitude increase in single-qubit coherence discussed above therefore enables a substantially improved two-qubit gate fidelity. This interaction generates an operation equivalent to a controlled phase gate up to single-qubit rotations. 

\begin{figure}
\includegraphics{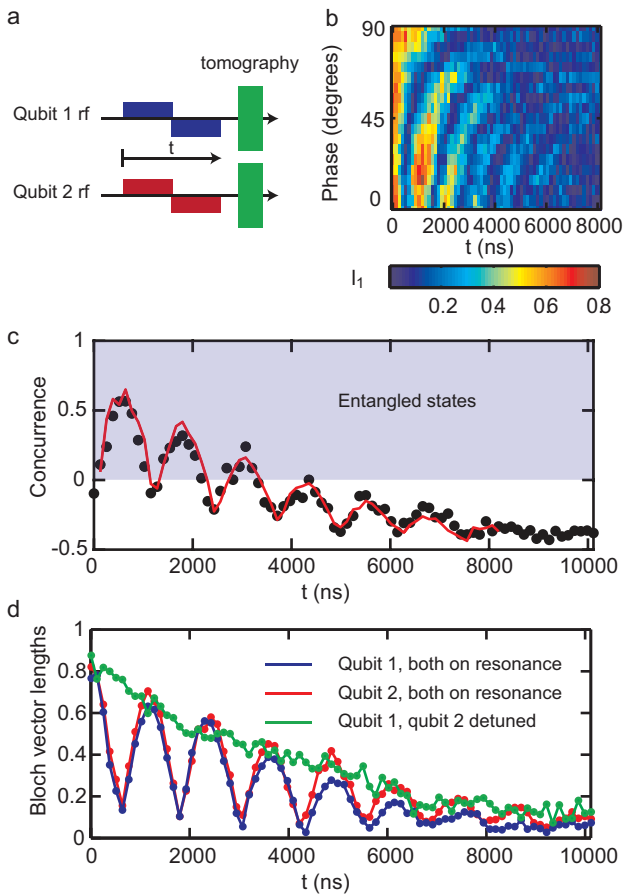}
\caption{\label{twoQubit} Entangling gate. (a) To entangle the qubits, we perform a simultaneous rotary echo on both for a varying total length of time $t$, followed by tomographic readout to reconstruct the two-qubit density matrix. (b) Bloch vector length for qubit 1, $l_1$, during the entangling gate as the phase between rf drives varies. Nodes in $l_1$ denote entanglement. The entanglement rate vanishes when the qubits are driven $90^{\circ}$ out of phase. (c) Concurrence vs. time for the entangling gate. Positive values of concurrence indicate entangled states. Negative values of concurrence indicate decoherence. The solid red line is a theoretical simulation taking into account hyperfine and low- and high-frequency charge noise (see Supplementary Information). (d) Single-qubit Bloch vector lengths $l_1$ and $l_2$ during joint evolution. When the qubits are detuned from each other, the interaction vanishes.}
\end{figure}

To entangle the qubits, we set $\W_1=\W_2=960$ MHz and perform a simultaneous rotary echo for varying lengths of time [Fig.~\ref{twoQubit}(a)], choosing the drive amplitude that maximizes the echo coherence time. Self-consistent two-qubit measurement and state tomography~\cite{Takahashi2013} (see Supplementary Information) reveal an oscillating concurrence~\cite{Hill1997} of the two-qubit state [Fig.~\ref{twoQubit}(c)]. The concurrence periodically reaches values above zero, indicating repeated entangling and disentangling of the qubits as the interaction time increases. Eventually, the concurrence saturates at a negative value, because both qubits have dephased. We have performed numerical simulations taking into account hyperfine noise and both low- and high-frequency charge noise (see Supplementary Information). The measured concurrence agrees with the simulation [Fig.~\ref{twoQubit}(c)]. As the concurrence reaches a local maximum, the length of the single-qubit Bloch vectors,  $l=\sqrt{\ta{\sigma_{x}}^2 +\ta{\sigma_{y}}^2+\ta{\sigma_{z}}^2}$, where $\ta{\cdots}$ indicates a single-qubit expectation value, approaches zero, as expected for entangled states [Fig.~\ref{twoQubit}(d)]. 

As equation~\ref{eq:Interaction} suggests, the interaction strength depends on the relative phase between the rf drives on each qubit. We demonstrate phase control of the two-qubit interaction by measuring the length of the Bloch vector of one qubit as we vary the relative phase between qubits [Fig.~\ref{twoQubit}(b)]. As the expected, the entangling rate reaches a maximum when the two qubits are driven in phase, and the entangling rate vanishes when the two qubits are out of phase. 

The two-qubit interaction also vanishes if $\W_1 \neq \W_2$. To demonstrate frequency control of the two-qubit gate, we turn off the dynamic nuclear polarization~\cite{Foletti2009} on qubit 2, effectively setting $\W_2 \approx 0$ MHz. However, all gate voltages during the entangling operation remain the same. Measuring the Bloch vector length of qubit 1 as a function of evolution time shows no oscillations, just a smooth decay [Fig.~\ref{twoQubit}(d)]. This indicates that no entanglement takes place, and hence that the interaction vanishes, when the two qubits are detuned from each other.

\begin{figure}
\includegraphics{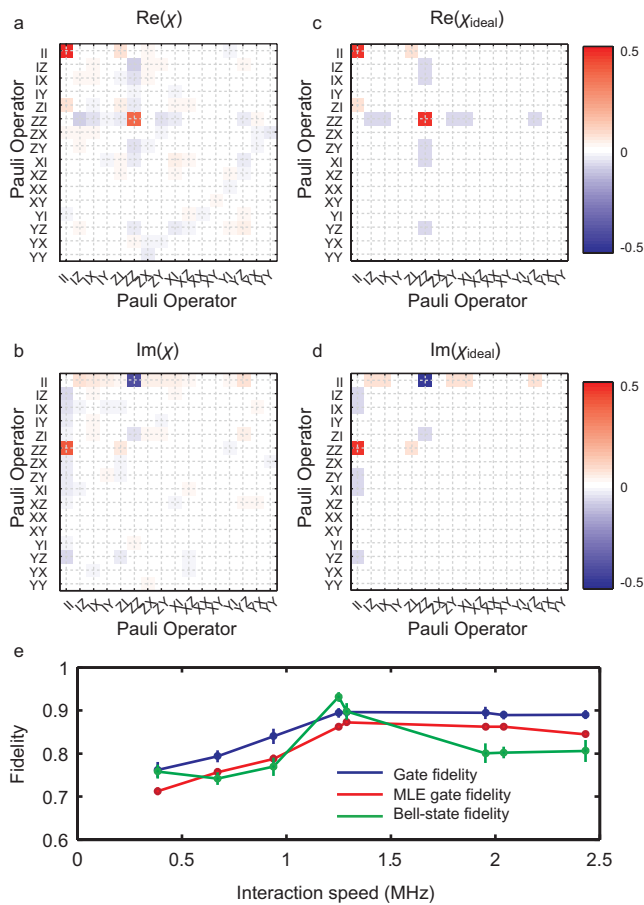}
\caption{\label{tomo} Process tomography for the two-qubit entangling gate. (a) Real component of the measured process matrix. (b) Imaginary component of the measured process matrix. (c) Real component of the ideal process matrix. (d) Imaginary component of the ideal process matrix. (e) Gate fidelity of the measured process matrix and most-likely completely positive process matrix and two-qubit Bell state fidelity as a function of interaction strength. Error bars are statistical errors.}
\end{figure}

To assess the gate fidelity, we perform self-consistent quantum process tomography~\cite{Chuang1997,Poyatos1997,Takahashi2013} on the two-qubit gate [Fig.~\ref{tomo}(a)-(d)], requiring 256 tomographic measurements of the two-qubit operation. We extract a maximum gate fidelity of $90 \pm 1 \%$ based on a measured tomographically complete set of input and output states (see Supplementary Information). The extracted process matrix $\chi$ has a few negative eigenvalues, which may result from partially mixed input states. Using a maximum likelihood estimation process to ensure a completely positive process matrix (see Supplementary Information), we extract a gate fidelity of $87 \pm 1\%$, which is consistent with the fidelity obtained by direct inversion.

Figure~\ref{tomo}(e) shows the maximum observed gate and Bell state fidelity as a function of interaction strength, which is varied by adjusting $J(\en)$ on each qubit. Similar to the case of single qubit coherence times, the gate fidelity drops at low interaction rates due to hyperfine noise. Gate fidelities are also expected to drop at fast interaction times due to charge noise, but we did not perform this experiment because our dynamic nuclear polarization feedback is not stable in this regime. An additional source of error at large interaction strengths is relaxation of the qubit states during initialization due to increased charge noise. We observe a maximum concurrence of $0.86 \pm 0.02$, corresponding to a Bell state fidelity of $93 \pm 1 \%$. Given that the observed  Bell state fidelities are equal to or slightly larger than the gate fidelities, it is likely that both decoherence and control errors play a role in overall gate fidelity. 

The maximum entangled state fidelity presented here represents a reduction in infidelity of about a factor of 4 over the previous entangling gate between singlet-triplet qubits~\cite{Shulman2012}, because the effects of charge noise are reduced when the magnetic gradient dominates the exchange interaction. This gate can be improved in the future by narrowing the hyperfine distribution through rapid Hamiltonian estimation~\cite{Shulman2014}, or by using spin qubits in nuclear-spin-free materials such as Si, where strong gradients can be established with micromagnets. We estimate that with laboratory frame coherence times of $1~\mu$s (instead of $\approx$ 100 ns here), rotating frame coherence times could increase by as much as 3-4 times. Longer coherence times such as these suggest that  two-qubit gate fidelities exceeding $99\%$, and fault-tolerant quantum computation using spins, are within reach.

\section{Methods}
The two double quantum dots are fabricated on a GaAs/AlGaAs hetereostructure with a two-dimensional electron gas located 91 nm below the surface. The two-dimensional electron gas density $n=1.5 \times 10^{11}$cm$^{-2}$ and mobility $\mu=2.5 \times 10^6$cm$^2/$Vs were measured at $T=4$K. Voltages applied to Au/Pd depletion gates define the double-dot potential. The qubits are cooled in a dilution refrigerator to a base temperature of approximately 20 mK. An external magnetic field B=0.7 T is applied in the plane of the semiconductor surface perpendicular to the axis of the double quantum dots. This orientation of the magnetic field ensures effective dynamic nuclear polarization~\cite{Nichol2015}.

\section{Acknowledgments}
We thank Stephen Bartlett and Andrew Doherty for valuable discussions. This research was funded by the United States Department of Defense, the Office of the Director of National Intelligence, Intelligence Advanced Research Projects Activity, and the Army Research Office grant W911NF-15-1-0203. S.P.H. was supported by the Department of Defense through the National Defense Science Engineering Graduate Fellowship Program. This work was performed in part at the Harvard University Center for Nanoscale Systems (CNS), a member of the National Nanotechnology Infrastructure Network (NNIN), which is supported by the National Science Foundation under NSF award No. ECS0335765.

\section{Author Contributions}
S.F., G.C.G., and M.J.M. grew and characterized the AlGaAs/GaAs heterostructure. S.P.H. fabricated the device. J.M.N. and L.A.O. performed the experiments. All authors discussed and analyzed the data and wrote the manuscript. A.Y. supervised the project.

\bibliography{rotating_frame}

\end{document}


\title{Supplementary Information for \\
		High-fidelity entangling gate for double-quantum-dot spin qubits}
	
	\author{John M. Nichol}
	\thanks{These authors contributed equally.}
	
	\author{Lucas A. Orona}
	\thanks{These authors contributed equally.}
	
	\author{Shannon P. Harvey}
	\affiliation{Department of Physics, Harvard University, Cambridge, MA, 02138, USA}
	
	\author{Saeed Fallahi}
	\affiliation{Department of Physics and Astronomy, Purdue University, West Lafayette, IN, 47907 USA}
	\affiliation{Birck Nanotechnology Center, Purdue University, West Lafayette, IN, 47907 USA}
	
	\author{Geoffrey C. Gardner}
	\affiliation{Birck Nanotechnology Center, Purdue University, West Lafayette, IN, 47907 USA}
	\affiliation{School of Materials Engineering, Purdue University, West Lafayette, IN, 47907 USA}
	
	\author{Michael J. Manfra}
	\affiliation{Department of Physics and Astronomy, Purdue University, West Lafayette, IN, 47907 USA}
	\affiliation{Birck Nanotechnology Center, Purdue University, West Lafayette, IN, 47907 USA}
	\affiliation{School of Materials Engineering, Purdue University, West Lafayette, IN, 47907 USA}
	\affiliation{School of Electrical and Computer Engineering, Purdue University, West Lafayette, IN, 47907 USA}
	
	\author{Amir Yacoby}
	\email{yacoby@physics.harvard.edu}
	
	\affiliation{Department of Physics, Harvard University, Cambridge, MA, 02138, USA}

\maketitle

\tableofcontents

\section{Qubit readout}\label{blah}
We operate each qubit with $\dbz\gg J(\en)$. The qubit eigenstates are approximately $\bra{\uparrow \downarrow}$ and $\bra{\downarrow \uparrow}$, where the left(right) arrow indicates the spin of the electron in the left(right) quantum dot in the (1,1) charge configuration, where each electron occupies its own quantum dot. These states can be read out by adiabatic charge transfer of both electrons into the right dot, where  $\bra{\uparrow \downarrow} \to \bra{T_0}$, and $\bra{\downarrow \uparrow} \to \bra{S}$. Pauli spin-blockade techniques are then used to distinguish $\bra{S}$ and $\bra{T_0}$. When $\dbz$ is large, rapid $\bra{T_0} \to \bra{S}$ relaxation occurs, diminishing readout contrast~\cite{Barthel2012}. To overcome this challenge, prior to adiabatic charge transfer, we adjust the electrochemical potential of the right quantum dot, such that an electron tunnels into the right dot. The other electron on the right dot then tunnels out, causing an $\bra{\uparrow \downarrow} \to \bra{\uparrow, S} \to \bra{T_+}$ transition. $\bra{\uparrow, S}$ is the lowest energy state with the (1,2) charge configuration with a polarized electron in the left dot, and a singlet in the right dot. $\bra{T_+}$ is the spin-polarized triplet. Following this sequence, we adiabatically transfer both electrons to the right dot and readout with Pauli spin-blockade. The key advantage of this technique is that $\bra{S} \to \bra{T_+}$ relaxation is much slower than $\bra{S} \to \bra{T_0}$ relaxation at large gradients. A detailed description of this readout procedure will be the subject of a future publication. Based on self-consistent measurement tomography, readout fidelities are approximately 75$\%$ (see section~\ref{section_tomo} below). This fidelity can be improved in the future by operating at larger magnetic field strengths.

\section{Qubit coherence time}\label{section_coherence}
We calculate the inhomogeneously broadened coherence time of driven Rabi oscillations. The amplitude of the total splitting in the rotating frame is, including noise, $\Wr=\sqrt{(j+\delta j)^2+\delta \W ^2}$. $j$ is the Rabi drive, $\delta j$ is the noise in the Rabi drive, and $\delta \W$ is the detuning noise.

Assuming that $j \gg \delta \W$, we have
\begin{eqnarray}
\delta \Wr=\delta j + \frac{1}{2}\frac{\delta \W^2}{j} \label{eq:noise}
\end{eqnarray}

$\delta j$ occurs because low-frequency charge noise modulates the value of $J'(\eps)$. Additionally, there are spectral components of the $\eps$-noise directly at $\w$. Set $\eps(t)=\eps_0 + \eps_1 \cos(\W t)+\delta \eps(t)$. Assume that the noise $\delta \eps(t) \ll \eps_0$.  Expanding $J(\eps)$, and keeping only terms upconverted by the modulation, we have
\begin{eqnarray}
	J(t)\approx & J(\eps_0) + \left( J'(\en) \eps_1  + J''(\en) \eps_1 \delta \eps(t) \right) \cos(\W t).
\end{eqnarray}
Assuming $\dbz \gg J(\en)$, we have $j=\frac{\eps_1}{2} J'(\en)$, and $\delta j(t) = \frac{1}{2} J''(\en) \eps_1 \delta \eps(t)$.

In GaAs qubits, noise in $\W$ arises primarily from fluctuations in $\dbz$. Note that $\delta \W$ renormalizes the mean value of $\ta{\Wr}=j + \frac{1}{2}\frac{\sigma_{\W} ^2}{j}$. We therefore compute 
\begin{eqnarray}
\ta{\frac{1}{4 j^2}\left( \delta \W^2-\sigma_{\W}^2 \right)^2}=&\frac{1}{4j^2}\left( \sigma_{\W}^4  -2 \sigma_{\W}^4 + 3 \sigma_{\W}^4 \right)\\
=&\frac{\sigma_{\W}^4 }{2j^2},
\end{eqnarray}
where we have made use of the fact that a fourth moment of a Gaussian variable is 3 times the standard deviation to the fourth power. Thus, in total, we have 
\begin{eqnarray}
\sigma_{\Wr}^2=\ta{\delta \Wr^2} &= \frac{\eps_1^2}{4} J''(\en)^2 \sigma_{\eps}^2+\frac{1}{\eps_1^2}\frac{\sigma_{\W}^4}{J'(\en)^2}.\label{eq:noiseTot}
\end{eqnarray}
At low drive strengths, the second term in equation ~\ref{eq:noiseTot}, which results from hyperfine noise, dominates. At large drive strengths, the first term, which results from charge-noise induced fluctuations in $j$, dominates. We therefore expect a minimum at intermediate values of $\eps_1$, where the inhomogeneously broadened coherence time $T_2^*=\frac{1}{\sqrt{2} \pi \sigma_{\Wr}}$ reaches a maximum.

\section{Two-qubit interaction}
The Hamiltonian in the lab frame is $H\approx \W_1 \sx \otimes I + \W_2 I \otimes \sx + J_{12}\sz\otimes\sz$, where for each qubit $\W \approx \dbz$. Transforming into the reference frame rotating around $\W \sx$, for each qubit,  $\sz \to \sz \cos(\W t+\phi)+\sy\sin(\W t + \phi)$. Therefore 
\begin{equation}
\begin{split}
H_{rot}&=J_{12}( \sz \otimes \sz \cos(\W_1t+\phi_1)\cos(\W_2t+\phi_2) + \\
& \sz \otimes \sy \cos(\W_1t+\phi_1)\sin(\W_2t+\phi_2) + \\
& \sy \otimes \sz \sin(\W_1t+\phi_1) \cos(\W_2t+\phi_2) + \\
& \sy \otimes \sy \sin(\W_1t+\phi_1)\sin(\W_2t+\phi_2)).
\end{split}
\end{equation}

$H_{rot}$ has a non-zero time averaged value only when $\W_1=\W_2$. In this case,
\begin{equation}
\begin{split}
\ta{H_{rot}}&=\frac{J_{12}}{2}( \sz \otimes \sz +\sy\otimes\sy)\cos(\phi_1-\phi_2)+\\
&\frac{J_{12}}{2}( \sz \otimes \sy - \sy\otimes\sz)\sin(\phi_1-\phi_2).
\end{split}
\end{equation}
If both qubits are driven in the rotating frame with different Rabi frequencies, they rotate around their $z$ axes at different rates, and all terms involving $\sy$ average to zero. Thus
\begin{equation}
\begin{split}
\ta{H_{rot}}&=\frac{J_{12}}{2} \sz \otimes \sz \cos(\phi_1-\phi_2).
\end{split}
\end{equation}

\section{Measurement, state, and process tomography}\label{section_tomo}
We perform self-consistent measurement- and state-tomography~\cite{Takahashi2013}, which requires at minimum state tomography on 4 known input states to reconstruct the positive operator valued measure (POVM) operators characterizing the three tomographic measurements per qubit. However, we can only initalize the qubit in its energy eigenbasis. Furthermore, the qubit state partially depolarizes during a 1.5 $\mu$s wait after initialization to let gate voltage stabilize. We load a singlet state in the (0,2) charge configuration with $>99\%$ probability. To assess the depolarization, we measure the amplitude of Rabi oscillations with and without the 1.5$~\mu$s wait, and attribute the loss in amplitude to depolarization. At large $J(\en)$, the amplitude diminishes by roughly $20 \%$, consistent with measurements of $T_1$. To generate the required number of states, we follow Ref.~\cite{Takahashi2013}
in evolving the prepared state under two evolution Hamiltonians (rf drives with different phases), performing state tomography at various times, and also fitting for the parameters describing the evolution Hamiltonians. In total, there are 11 unknowns for single-qubit tomography: 2 parameters describing projection fidelities, 9 parameters describing the three measurment axes (3 for each) and 6 parameters describing the evolution Hamiltonians (3 for each). We perform state tomography for 48 different states: 16 measurements of the prepared state, 16 at different times for one rf drive, and 16 at different times for the other rf drive. All data are fitted simultaneously to calibrate the tomography. Calibrations are consistent from run to run, and typical readout fidelities are $\approx 75 \%$. To calibrate measurements for two-qubit tomography, we perform single-qubit calibrations for each qubit on all pairs of single qubit measurements. Based on our measurements, we find the most likely physical density matrix using the expressions derived in Ref.~\cite{Smolin2012}. 

Process tomography is performed through state tomography on 16 input and 16 output states for our two-qubit gate. Using the measured input and output states, we invert the equation for the process matrix: $E(\rho)=\sum_{m,n=1}^{16} \chi_{mn}B_n \rho B_m^{\dagger}$. Here $E$ is the map representing our two-qubit gate, $\chi$ is the process matrix, and $B_m=\sigma_i \otimes \sigma_j$ are combinations of the Pauli operators. Our inverted process matrix has some negative eigenvalues, which may result from our mixed input states. We can constrain the process matrix to be completely positive and trace-preserving, by constraining the Choi matrix to be positive semidefinite and requiring that the partial trace over the qubit equal the identity~\cite{Chow2012}. The maximum-likelihood algorithm is implemented with the Matlab CVX library (www.cvxr.com).

To find the ideal process matrix, $\chi_{ideal}$, we start with the process matrix generated by the interaction Hamiltonian $H_{int}=\sz \otimes \sz$, and search through all single-qubit rotations to find the highest fidelity, given by Tr$(\chi_{ideal}\chi )$. The largest single-qubit rotations occur around the rf drive axis, because of rise-time effects in the coaxial cables in our cryostat. Bell state fidelities are found by searching through all single-qubit rotations for the Bell state with the largest overlap with the measured state.

Uncertainties in the state and gate fidelities are obtained using the measured experimental uncertainties in our data. We add Gaussian distributed noise to the data with the measured standard deviation and reconstruct noisy density and process matrices and corresponding noisy state and gate fidelities. The quoted uncertainties are the standard deviation of 128 different noisy fidelities. The mean of the noisy fidelities generated in this way agrees with the measured fidelity.

\section{Simulation}
To generate the simulated curve in Fig. 2(c) in the main text, we numerically integrated the Schr\"odinger equation using the laboratory frame Hamiltonian $H=J_1\left(\eps_1(t)\right)\sz\otimes I+\dbz\sx \otimes I +J_2(\eps_2(t))I \otimes \sz+\dbz I\otimes \sx +J_{12}\sz \otimes \sz$ with time-varying voltages $\eps_1(t)$ and $\eps_2(t)$ for each qubit and computed the concurrence of the resulting states. We used $\W_1 = \W_2 = 960$ MHz and adjusted the interaction strength to match the observed entanglement frequency. We used the measured functional form of $J(\eps)$ for each qubit. We assumed independent fluctuations in $\dbz$ corresponding to laboratory frame coherence times of 150 ns, independent low-frequency charge fluctuations with standard deviation 8 $\mu$V relative to the gates, and independent high frequency charge noise with power spectrum $S(f)\propto f^{-0.7}$ with a magnitude of 0.9 nV/$\sqrt{\textnormal{Hz}}$ at 1 MHz up to $f=1$ GHz and 0 otherwise for each qubit. These values for charge noise are consistent with previous measurements in GaAs singlet-triplet qubits~\cite{Dial2013}.

\bibliography{rotating_frame_SI}